\documentclass[aps,prl,twocolumn,superscriptaddress,groupedaddress]{revtex4}

\usepackage{graphicx}  
\usepackage{dcolumn}   
\usepackage{bm}        
\usepackage{amssymb}   

\begin{document}

\title{Topology-Scaling Identification of Layered Solids and Stable Exfoliated 2D Materials}

\author{Michael Ashton}
\affiliation{
 Department of Materials Science and Engineering,\\
 University of Florida, Gainesville, FL 32611-6400
}

\author{Joshua Paul}
\affiliation{
 Department of Materials Science and Engineering,\\
 University of Florida, Gainesville, FL 32611-6400
}

\author{Susan B. Sinnott}
\affiliation{
 Department of Materials Science and Engineering,\\
 The Pennsylvania State University, University Park, PA 16801-7003
}

\author{Richard G. Hennig}
\affiliation{
 Department of Materials Science and Engineering,\\
 University of Florida, Gainesville, FL 32611-6400
}

\date{\today}

\begin{abstract}
  The Materials Project crystal structure database has been searched
  for materials possessing layered motifs in their crystal structures
  using a topology-scaling algorithm. The algorithm identifies and
  measures the sizes of bonded atomic clusters in a structure's unit
  cell, and determines their scaling with cell size. The search
  yielded 826 stable layered materials that are considered as
  candidates for the formation of two-dimensional monolayers via
  exfoliation. Density-functional theory was used to calculate the
  exfoliation energy of each material and 680 monolayers emerge with
  exfoliation energies below those of already-existent two-dimensional
  materials. The crystal structures of these two-dimensional materials
  provide templates for future theoretical searches of stable
  two-dimensional materials. The optimized structures and other
  calculated data for all 826 monolayers are provided at
  \url{https://materialsweb.org}.
\end{abstract}

\maketitle

The combination of modern computational tools and the growing number
of available crystal structure databases with high-throughput
interfaces have accelerated recent efforts to map the materials
genome. One of the most recently discovered branches of the materials
genome is the class of two-dimensional (2D) materials, which generally have
properties that are markedly different from their three-dimensional
counterparts. The canonical example is the graphite/graphene system,
but monolayers have been exfoliated from many other layered compounds
as well \cite{joensen1986single, lin2009soluble,
altuntasoglu2010syntheses, coleman2011two}. Stable 2D materials can also be
obtained by deposition \cite{lee2012synthesis, cong2014synthesis, li2011large,
singh2014prediction} or chemical exfoliation \cite{naguib2011two,
ashton2016predicted}. Because the contribution of interlayer interactions to
these materials’ free energies is typically quite small, the existence of a
mechanically exfoliable bulk precursor generally indicates the relative
stability of a free-standing single layer, regardless of how it is synthesized.

In an effort to discover novel 2D materials, two recent studies
searched the inorganic crystal structure database (ICSD) for compounds
with large interlayer spacings, which are characteristic of weak
interlayer bonding that could be overcome by mechanical exfoliation
\cite{bjorkman2012van, lebegue2013two}. They used the intuitive
criteria of a low packing fraction based on the covalent radii of the
atoms and an interlayer gap larger than the sum of the covalent radii
of atoms at the layers' surfaces along the \textit{c}-axis to identify
layered compounds in the ICSD. They discovered almost 100 layered
phases, nearly half of which had monolayers that had not been the
subject of any prior publications.

Here, we extend their method to identify a large number of layered
compounds that were missed using the packing factor and
\textit{c}-axis interlayer gap criteria. We further add the constraint
that a bulk material must be thermodynamically stable to be of
interest during our search. Therefore, we use the Materials Project
(MP) database \cite{Jain2013}, an online repository of crystallographic
and thermodynamic data for over 65,000 compounds calculated with
density-functional theory (DFT).

\begin{figure}
  \includegraphics[width=8cm]{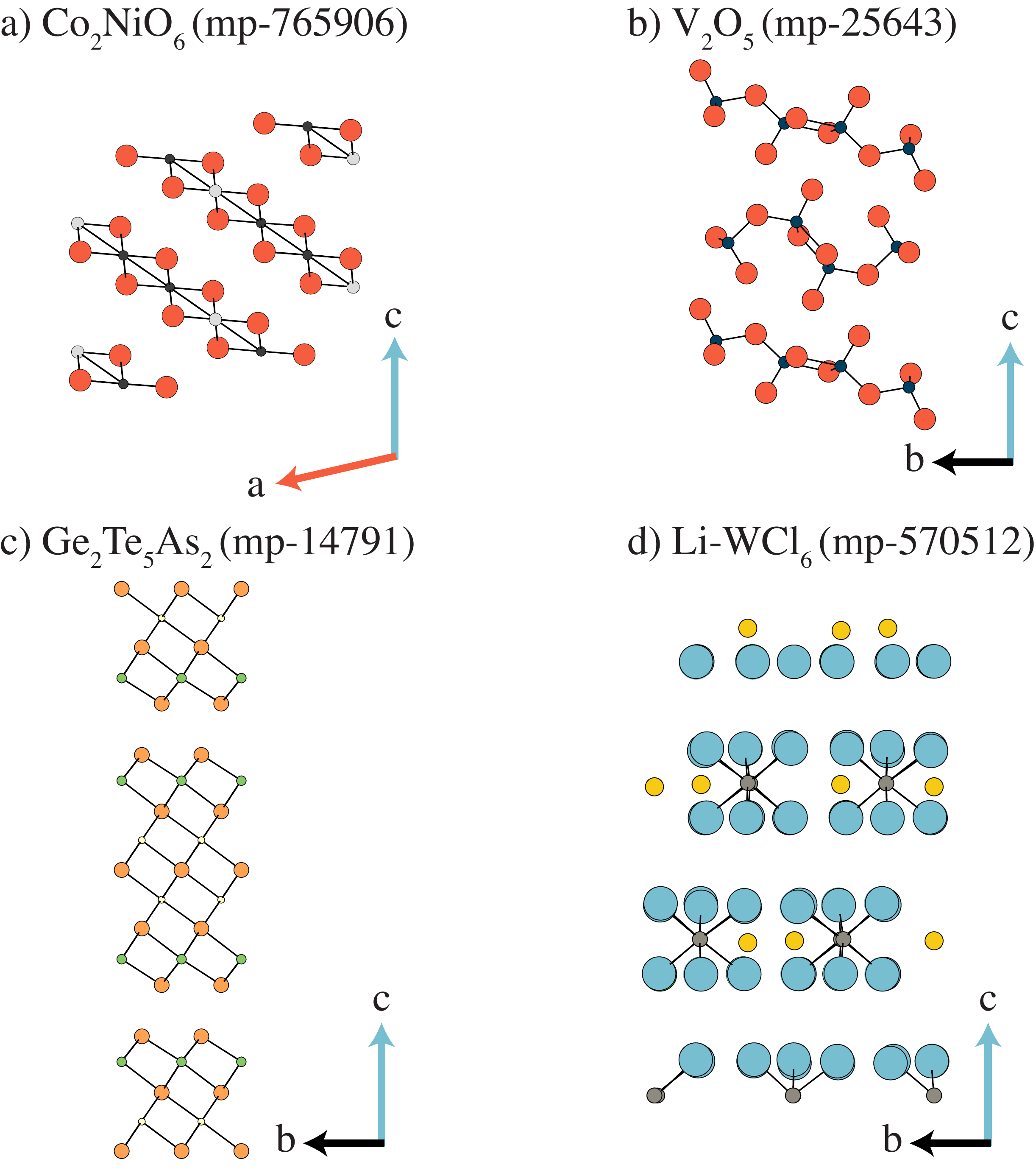}
  \caption{\label{fig:exceptions} (Color online) Examples of structure types
    that challenge the search for layered materials by exhibiting a)
    layers along an axis other than the \textit{c}-axis, b) corrugated
    layers, c) thick layers, and d) molecular (non-bonded)
    layers. Each of these examples is correctly classified with the
    present method.}
\end{figure}

Our algorithm is designed to correctly identify additional layered
materials by handling four different cases that are illustrated in
Fig.~\ref{fig:exceptions}: (a) materials whose layers are not
perpendicular to unit-cell axes or parallel to unit cell surfaces,
(b) materials with corrugated layers that therefore lack a planar
interlayer spacing, (c) materials with very thick layers that exceed
the packing factor normally observed for layered materials, and (d)
materials composed of molecular species that have gaps along multiple
axes and are often identified as false positives.

\begin{figure}
  \includegraphics[width=6cm]{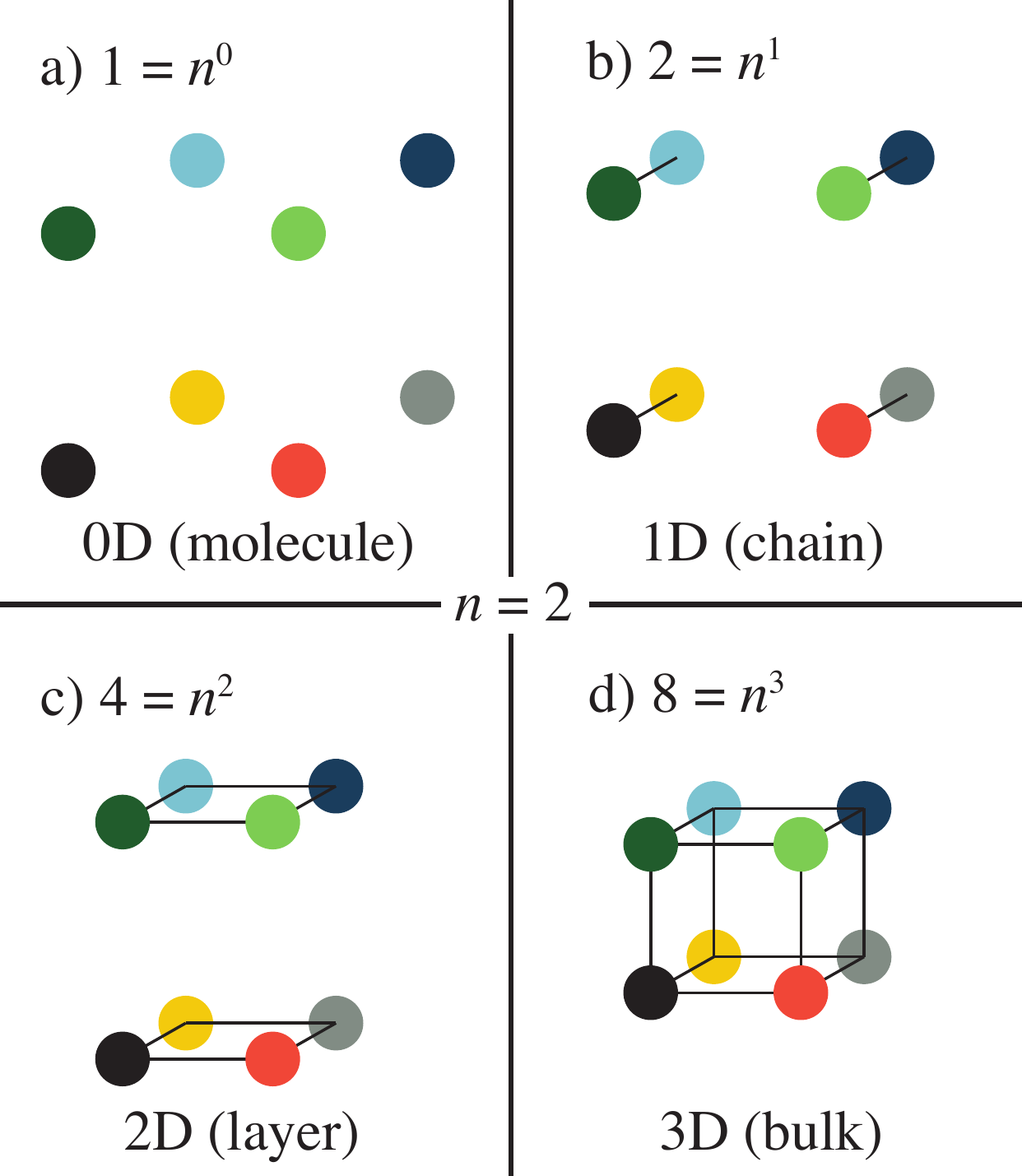}
  \caption{\label{fig:algorithm} (Color online) Schematic of the
    topology-scaling algorithm (TSA). The black circle (bottom left in
    each frame) represents a cluster of bonded atoms in a given
    crystal structure's unit cell. The colored circles represent the
    seven periodic images of the same cluster for a
    2$\times$2$\times$2 supercell of the original cell. The periodic
    images are either bonded to one another in a) zero, b) one, c)
    two, or d) three dimensions, defining the dimensionality of the
    structural motifs.}
\end{figure}

Each of these cases is correctly classified using our topology-scaling
algorithm (TSA) to identify layered compounds. The first step in the
TSA is to isolate bonded clusters of atoms in the structure, where a
bond is defined as an overlap in the covalent radii of two neighboring
atoms plus a small tolerance. If all atoms in the structure are in the
same cluster, the structure is classified as a conventional bulk
compound. If not, we count the number of atoms in a single cluster,
create an $n\times n \times n$ supercell of the original structure,
and group all atoms into bonded clusters again. The scaling of the
cluster size with supercell size, $n$, determines the dimensionality
of the structure as illustrated in Fig.~\ref{fig:algorithm}. If the
number of atoms in the original cluster does not change with supercell
size (zeroth order scaling), the cluster is an isolated molecule or
atom. If the number of atoms scales linearly with $n$, the cluster is
a one-dimensional chain. If it scales as $n^2$, it is a true layered
solid. If it scales as $n^3$, it is most likely an intercalated solid
(\textit{e.g}. a lithiated zeolite) composed of a three-dimensional bonded
network structure with intercalated atoms or molecules.

Here, we use the TSA to identify layered solids, but its ability to
simultaneously identify bonded networks of any dimension is what
allows it to systematically classify structures in large materials
databases. Correctly distinguishing between molecular, intercalated,
and layered solids is crucial, since there are large numbers of
molecular and intercalated structures in most materials databases.

The strength of the TSA is that it discovers structural motifs that
are separated from each other by distances larger than the bond length
of atoms within the motifs. The search for bonded clusters employs a
range of bond-length tolerances between 100\% and 135\% of the summed
covalent radii \cite{cordero2008covalent}. A few compounds are only
identified as layered for a small range of tolerances; these are
visually examined.

Our algorithm identifies 1560 layered materials, 509 of which have
zero distance to their respective thermodynamic convex hulls, and
hence are predicted to be stable compounds. Another 590 materials are metastable
by less than 50~meV/atom, 206 by between 50 and 100~meV/atom, and 255 are
unstable by more than 100~meV/atom. Here, we focus on the stable and
metastable materials with distances to the hull of less than
50~meV/atom. Additionally, several of the 1560 layered materials are
simply different stacking sequences of the same monolayer. These
duplicates are filtered out using symmetry analysis resulting in
826 unique monolayers for further investigation.

\begin{figure*}
\includegraphics[width=16cm]{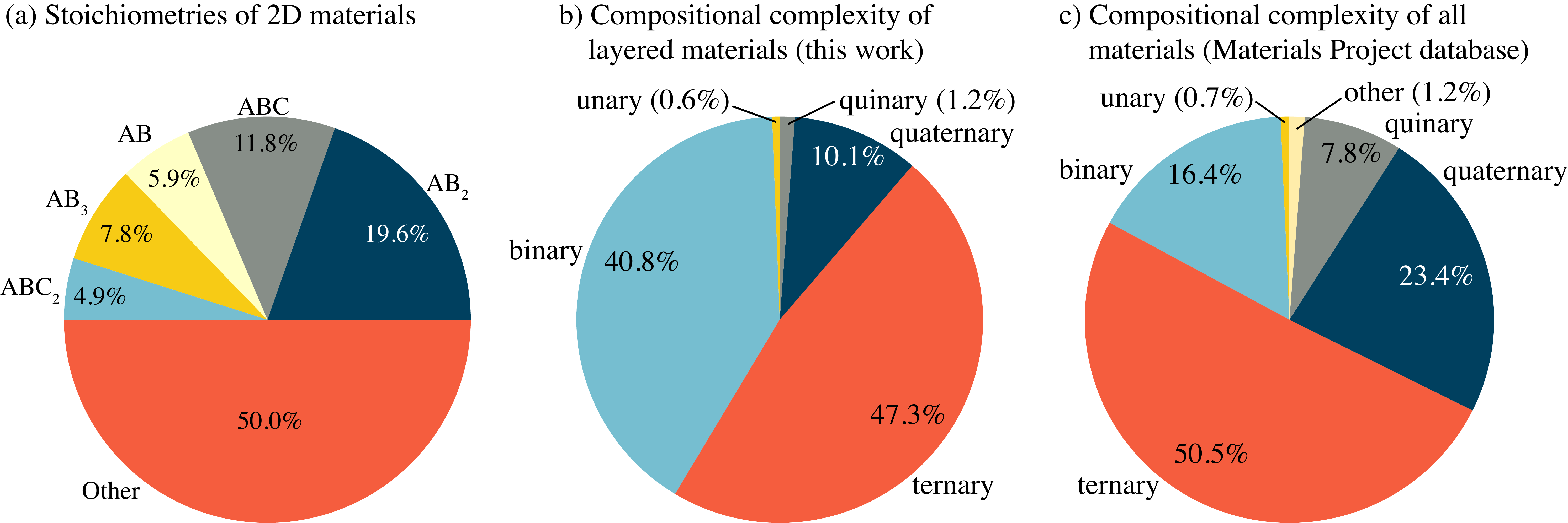}
\caption{\label{fig:complexities} (Color online) Distribution of (a)
  stoichiometries of the 826 layered compounds, and comparison of the
  compositional complexity among (b) the stable layered materials
  identified by this work and (c) all materials in the Materials
  Project database. The top 5 stoichiometries (ABC, AB$_2$, AB,
  AB$_3$, and ABC$_2$) represent half of all compounds. In general,
  the relative abundance of a given stoichiometry scales inversely
  with the formula's complexity. We observe that the percentage of
  binary compounds among layered materials is significantly higher
  than among all materials, suggesting that binary compounds (one
  cation and one anion) are particularly conducive to creating
  interlayer dispersion interactions.}
\end{figure*}

These 826 materials can be grouped according to their stoichiometric
ratios. Fig.~\ref{fig:complexities}(a)
shows that 50\% of the layered materials are represented by just five
stoichiometries. These five -- AB$_2$, ABC, AB, AB$_3$, and ABC$_2$,
in decreasing order of frequency -- mostly reflect the abundance of
known 2D materials with simple cation/anion stoichiometries. However,
the large variety of 102 unique stoichiometries
indicates that the family of potentially exfoliative layered compounds
is much more diverse than the simple compositions typically
considered as 2D material candidates.

Figure~\ref{fig:complexities}(b) and (c) compare the percentages of
unary, binary, ternary, {\it etc.} stable layered compounds in the MP
database with the percentages of all stable (distance to hull $<$
50~meV/atom) compounds in the MP database. Binary, ternary, and
quaternary compounds comprise 98.2\% of the stable layered
compounds. The percentages of unary and ternary compounds among
layered materials are very close to their percentages among all
materials in the MP database. However, binary compositions are clearly
overrepresented among layered materials, while quaternary and quinary
compositions are underrepresented. This indicates that compounds of
two or three species can more easily form low-energy structures that
exhibit dispersion-bound layers.

\begin{figure}
\includegraphics[width=\columnwidth]{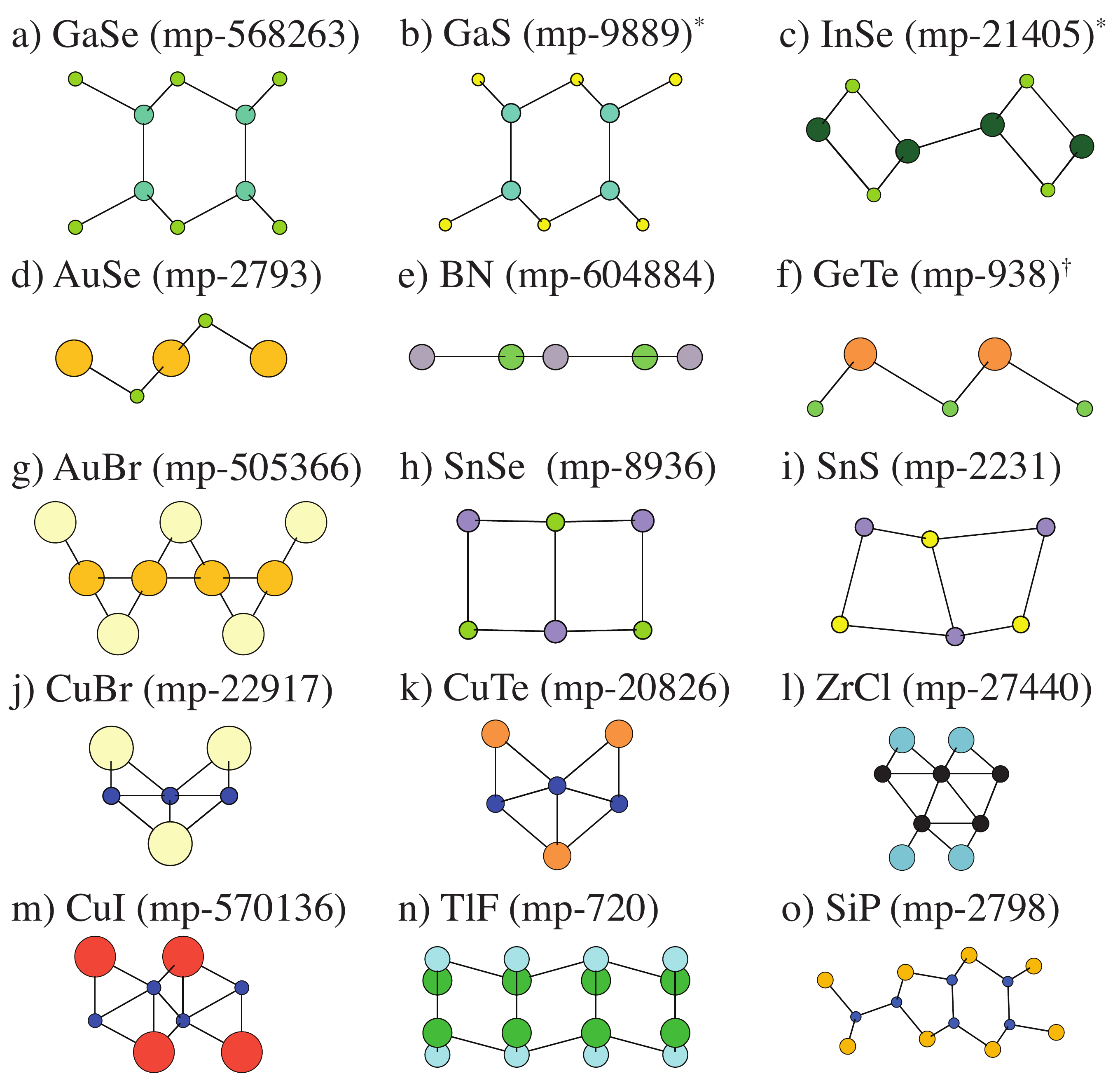}
\footnotetext{$^{\ast}$ Not minimum energy structure for any compound we observed.}
\footnotetext{$^{\dagger}$ Exfoliation energy over 200 meV/atom.}
\caption{\label{fig:ab-templates} (Color online) Side views of the 15
  unique structural templates identified for monolayers of AB
  stoichiometry, labeled according to prototypical compounds
  possessing that crystal structure. The symmetries of structures a)
  and b) are related by a broken inversion symmetry, in the same way
  that the well-known 1T and 2H monolayer structures are
  related. Other structural pairs, such as e), f) and h), i) and j), k), are
  related to one another by simple buckling distortions.}
\end{figure}

These materials can be further classified according to their crystal
structures, which has implications for future searches
for 2D materials based on chemical substitutions and genetic
algorithms~\cite{singh2014computational, revard2016grand}. The crystal
structures represented by each stoichiometry can be used as templates
for these searches, which require viable crystal structures as the
initial parent structures to seed the algorithm. The use of a larger
number of seed structures increases the likelihood that a given search
will identify the global minimum of a composition's high-dimensional
phase space. The complete set of the structural templates for all stoichiometries
identified by our algorithm is available online in our database at
\url{https://materialsweb.org}.

An example of structural templates is given in Fig.~\ref{fig:ab-templates},
which shows all 15 unique crystal structures identified for the AB
stoichiometry. These 15 structures can either augment or stand
alone as the parent generation in future searches for 2D materials of
AB composition. These various structures will likely possess a wide
variety of properties, even for the same A and B elements. For
example, buckled structures, such as structures (i) and (m) in
Fig.~\ref{fig:ab-templates}, are uniquely interesting because their
broken inversion symmetry could lead to
piezoelectricity~\cite{duerloo2012intrinsic, zhuang2014computational,
  wu2014piezoelectricity, blonsky2015ab, fei2015giant} or, given
sufficiently large spin-orbit coupling, to Rashba spin
splitting~\cite{picozzi2014ferroelectric, zhuang2015rashba}. For
lubricating applications, flat monolayers like structures (f) and (h)
may exhibit lower friction coefficients than strongly buckled
monolayers, like structures (c) and (o). Top and side views of these
2D structures are provided in the Supplemental
Material~\cite{Supplement}.

We re-optimize the structures of all 826 stable layered materials and
the single monolayers from each structure with DFT using VASP
\cite{kresse1993ab, kresse1994ab, kresse1996efficiency,
  kresse1996efficient}. We employ the dispersion-corrected vdW-optB88
exchange-correlation functional \cite{dion2004van, roman2009efficient,
  klimevs2009chemical, klimevs2011van} to accurately account for
interlayer dispersion interactions. Comparison with more accurate 
and computationally demanding random-phase approximation (RPA)
calculations \cite{bjorkman2012van} show that this functional
reproduces relative trends in exfoliation energy and predicts the
interlayer interactions within 30\% of the RPA. We compare the
resulting exfoliation energy of the single monolayer materials, given
by the energy difference between the monolayer and the bulk solid,
with that of free-standing monolayers that have previously been
synthesized. As an upper bound, 2D SnSe has been synthesized
\cite{li2013single, ma2014growth} and has an exfoliation energy close
to 150 meV/atom. The most promising 2D material candidates will have
much lower exfoliation energies, but monolayers with exfoliation
energies below 150 meV/atom may be feasible for synthesis
\cite{singh2015computational, revard2016grand}.

\begin{figure}[t]
\includegraphics[width=7cm]{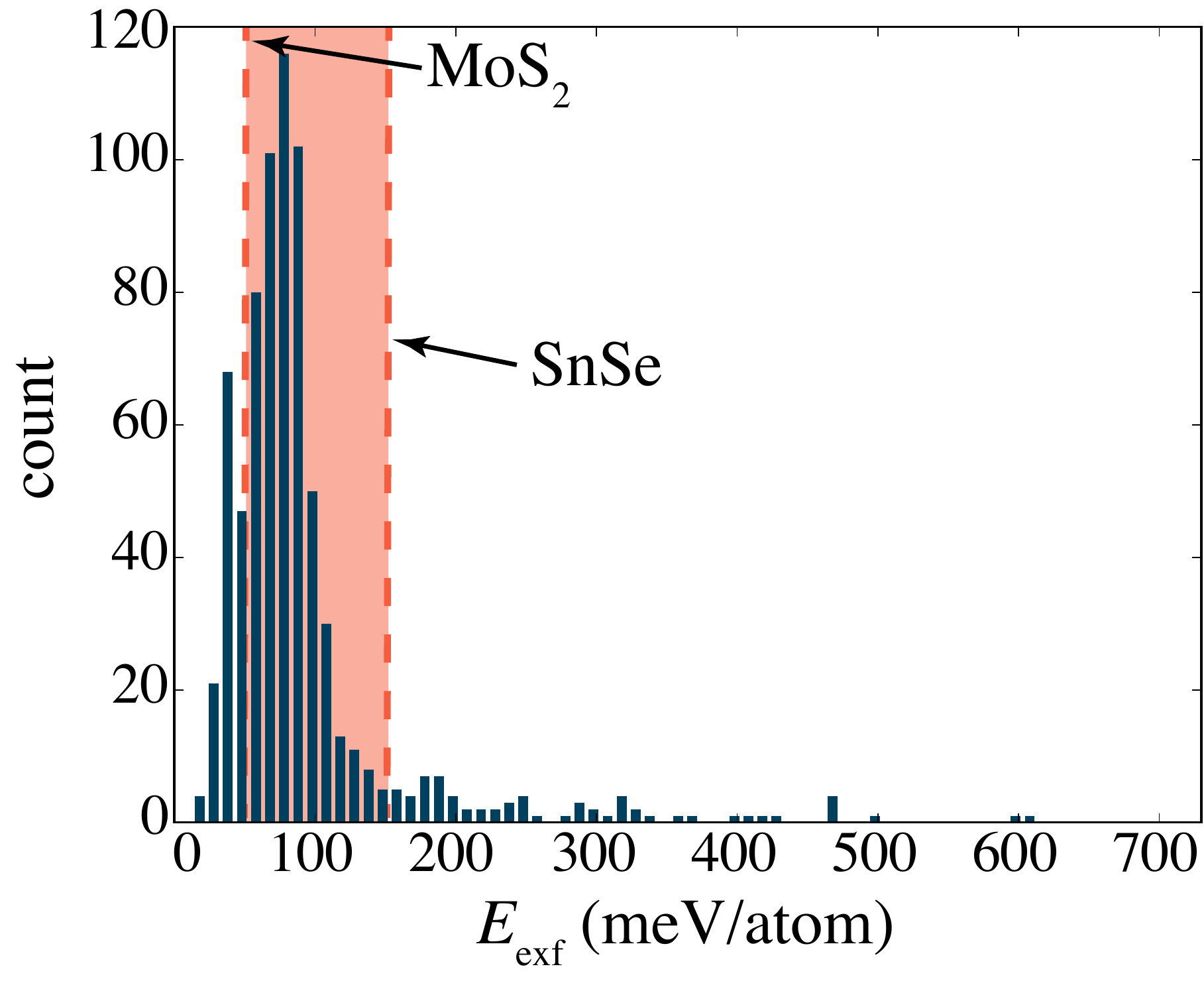}
\caption{\label{fig:exfoliation-energies} (Color online) Histogram of
  calculated exfoliation energies for all 826 layered materials
  compared to the range of calculated exfoliation energies for already
  synthesized 2D materials. Because of their relatively weak
  interlayer forces, most compounds found in our search exhibit low
  exfoliation energies ($<100$~meV/atom), indicating the ease of
  exfoliation.}
\end{figure}

Fig.~\ref{fig:exfoliation-energies} shows the distribution of
calculated exfoliation energies for all 826 compounds with respect to
the exfoliation energies of known 2D materials. A large majority of
680 compounds have exfoliation energies below 150~meV/atom, and most
(612) have exfoliation energies below 100 meV/atom.

\begin{figure}
  \includegraphics[width=8.5cm]{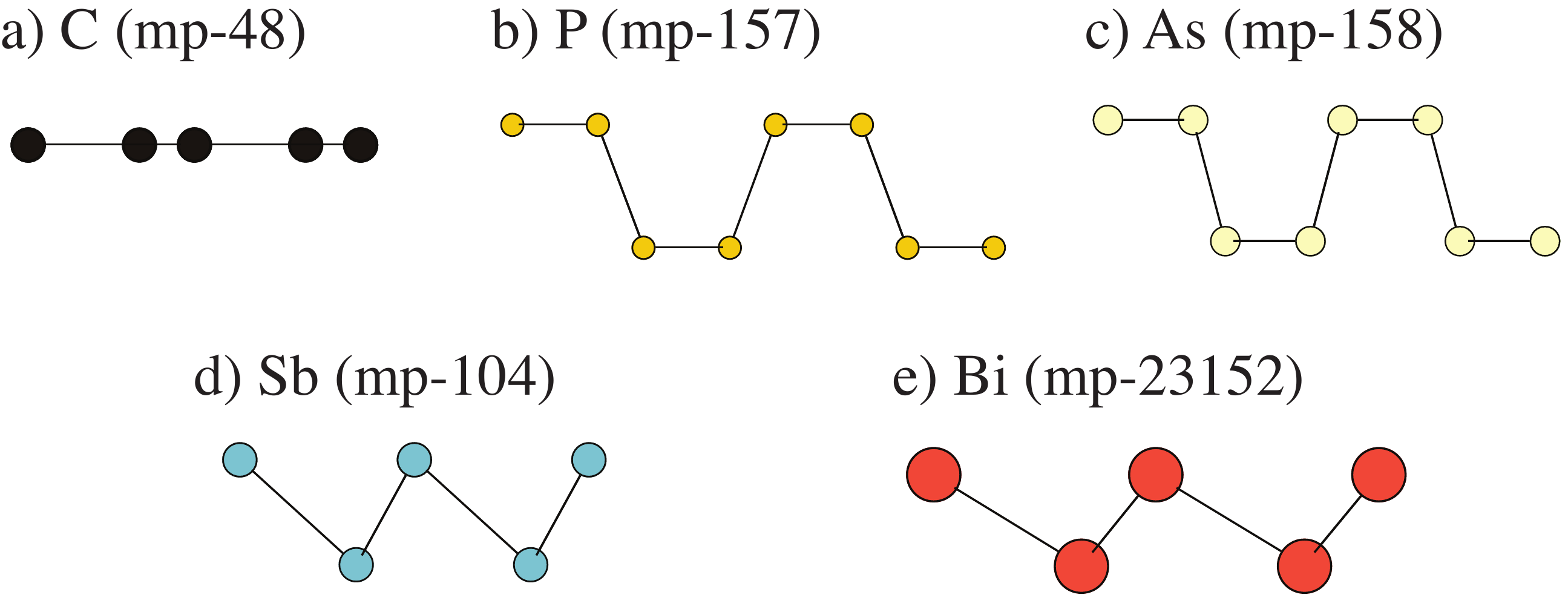}
  \caption{Side views of monolayers for the five elemental layered materials in
    the Materials Project database.}
  \label{fig:unaries}
\end{figure}

We find that C, P, As, Sb, and Bi are the only elements that have pure
layered forms. Of those five, Sb and Bi share the same structure, as
do P and As, leading to the three unique unary layered crystal
structures shown in Fig.~\ref{fig:unaries}.  Graphene (C), phosphorene
(P), arsenene (As), and antimonene (Sb) have all been reported as 2D
materials experimentally and/or theoretically
\cite{novoselov2004electric, liu2014phosphorene, zhang2015atomically,
  ares2016mechanical}, but very little work has been done to isolate
or characterize bismene (Bi) nanosheets. The exfoliation energy of
bismene is 273~meV/atom, similar to that of 2D Sb (236 meV/atom). Both
are above the 150 meV/atom threshold of exfoliation energies for
experimentally existent materials, but still much more stable than
silicene \cite{vogt2012silicene, zhuang2012electronic}, which has been
stabilized on substrates. Thus, 2D bismene, if not stable as a
free-standing monolayer, may be stabilized on a suitable substrate
\cite{singh2014ab, singh2014synthesis}. Top and side views of these 2D
structures are provided in the Supplemental
Material~\cite{Supplement}.

To illustrate the variety of properties accessible among the 826 2D
materials we have identified, 182 are predicted to exhibit magnetic
moments larger than 1~$\mu_{B}$/unit cell and 519 display band gaps at
the PBE level. Only ferromagnetic configurations were considered; more
detailed analysis will be required to determine which magnetic 2D
materials have antiferromagnetic or more complex orderings. CaClF and
MgCl$_2$ have the largest band gaps at the PBE level of both 6~eV. Their
inexpensive constituent elements and relatively substrate-agnostic
interlayer interactions make CaClF and MgCl$_2$ particularly
interesting candidates for thin transparent dielectrics in electronic
device technologies.

\begin{figure}
\includegraphics[width=8cm]{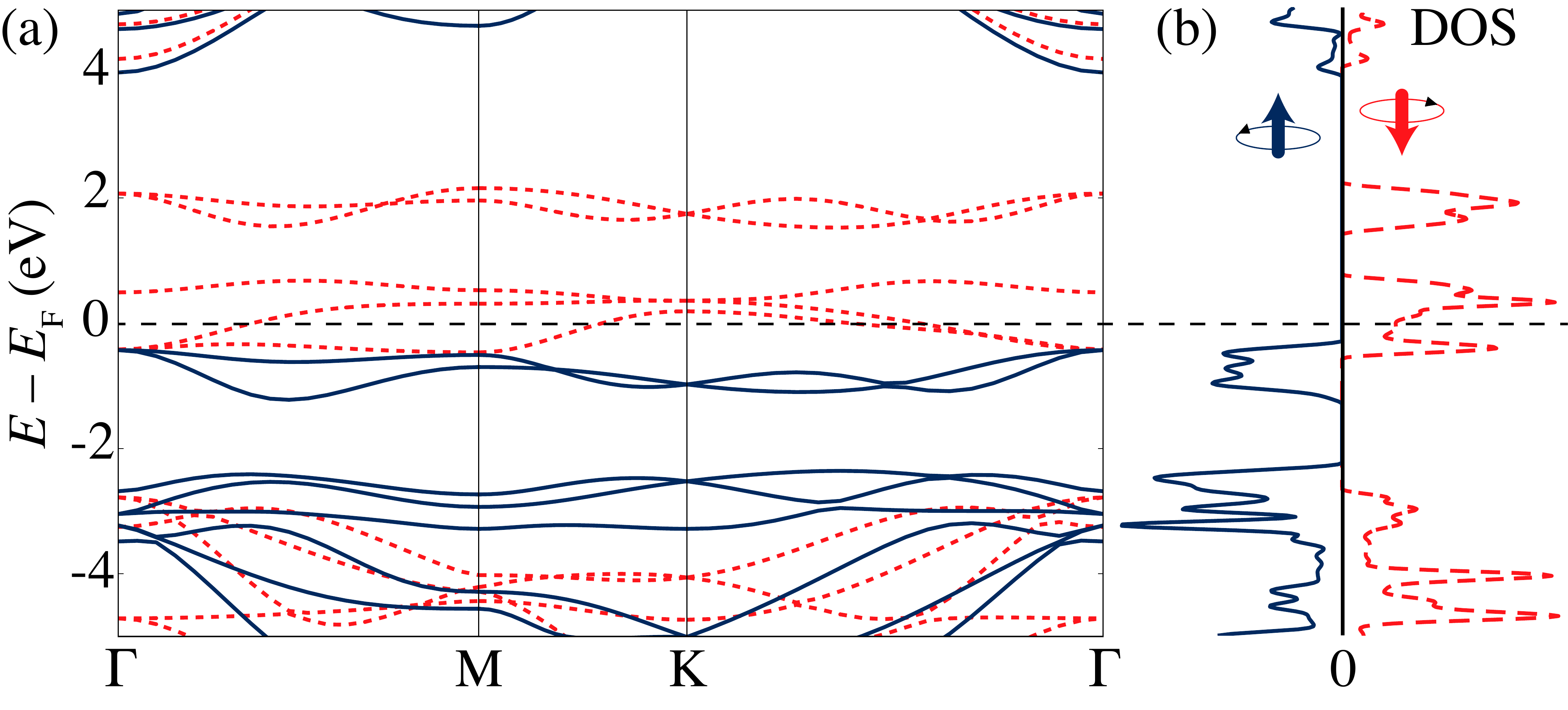}
\caption{(Color online) (a) Half-metallic band structure and (b) spin-polarized
  density of states for 2D FeCl$_2$. The minority spin state is
  metallic (dashed red) and the majority state (solid blue) has a band
  gap of 4.4~eV.}
\label{fig:HM-bands}
\end{figure}

Thirty of the magnetic 2D materials exhibit half-metallic character; that is,
exactly one of their spin channels is metallic. The band structure of FeCl$_2$,
an example half-metallic material, is provided in Figure~\ref{fig:HM-bands}. This
material's minority spin state is metallic, while its majority state is
insulating with a relatively large gap of 4.4 eV. Electrical currents in 2D
FeCl$_2$ and other half-metals will in principle be completely spin-polarized
\cite{de1983new, park1998direct, fang2002spin}, enabling the giant
magnetoresistive effect that powers many spintronic devices \cite{baibich1988giant}.
These materials are therefore of significant technological interest, as
their dispersive interlayer interactions make them uniquely adaptable to
vertically stacked heterostructures in, for example, spin valves
\cite{dieny1991magnetotransport, hill2006graphene} and Josephson junctions
\cite{eschrig2003theory}.

To view interactive and downloadable information for all
structures, including exfoliation energies, input files used for
calculations, band structures, and calculated Pourbaix diagrams, the
reader is again referred to our online database at
\url{https://materialsweb.org}.

In summary, we developed a general topology-based algorithm that
classifies crystal structures by the dimensionality of their
structural motifs and applied it to predict the stability of more than
600 potential 2D materials, including bismene, which may be the most
stable elemental 2D material left to be synthesized. The synthesis and
further characterization of these materials could, in turn, unearth a
commensurate wealth of materials properties.  Additionally, the
monolayer structures identified in this search can serve as viable
structural templates for theoretical searches for 2D materials, such
as genetic algorithm and chemical substitution searches, that may lead
to the discovery of many more stable 2D materials.

The authors thank Dr.\ Vin Crespi for valuable discussions. M.A.\ and
S.B.S.\ were supported by the National Science Foundation under grant
DMR-1307840, and J.P.\ and R.G.H.\ were supported under grants
DMR-1542776 and PHY-1549132, the Center for Bright Beams. All
calculations were performed using the HiPerGator supercomputer at the
University of Florida's High Performance Computing Center.


\end{document}


\title{Supplementary Information: Topology-Scaling Identification of Layered Solids and Stable Exfoliated 2D Materials}

\author{Michael Ashton}
\affiliation{
 Department of Materials Science and Engineering,\\
 University of Florida, Gainesville, FL 32611-6400
}

\author{Joshua Paul}
\affiliation{
 Department of Materials Science and Engineering,\\
 University of Florida, Gainesville, FL 32611-6400
}

\author{Susan B. Sinnott}
\affiliation{
 Department of Materials Science and Engineering,\\
 The Pennsylvania State University, University Park, PA 16801-7003
}

\author{Richard G. Hennig}
\affiliation{
 Department of Materials Science and Engineering,\\
 University of Florida, Gainesville, FL 32611-6400
}

\date{\today}

\maketitle

\renewcommand{\thefigure}{S\arabic{figure}}

\begin{figure}
  \includegraphics[width=\textwidth]{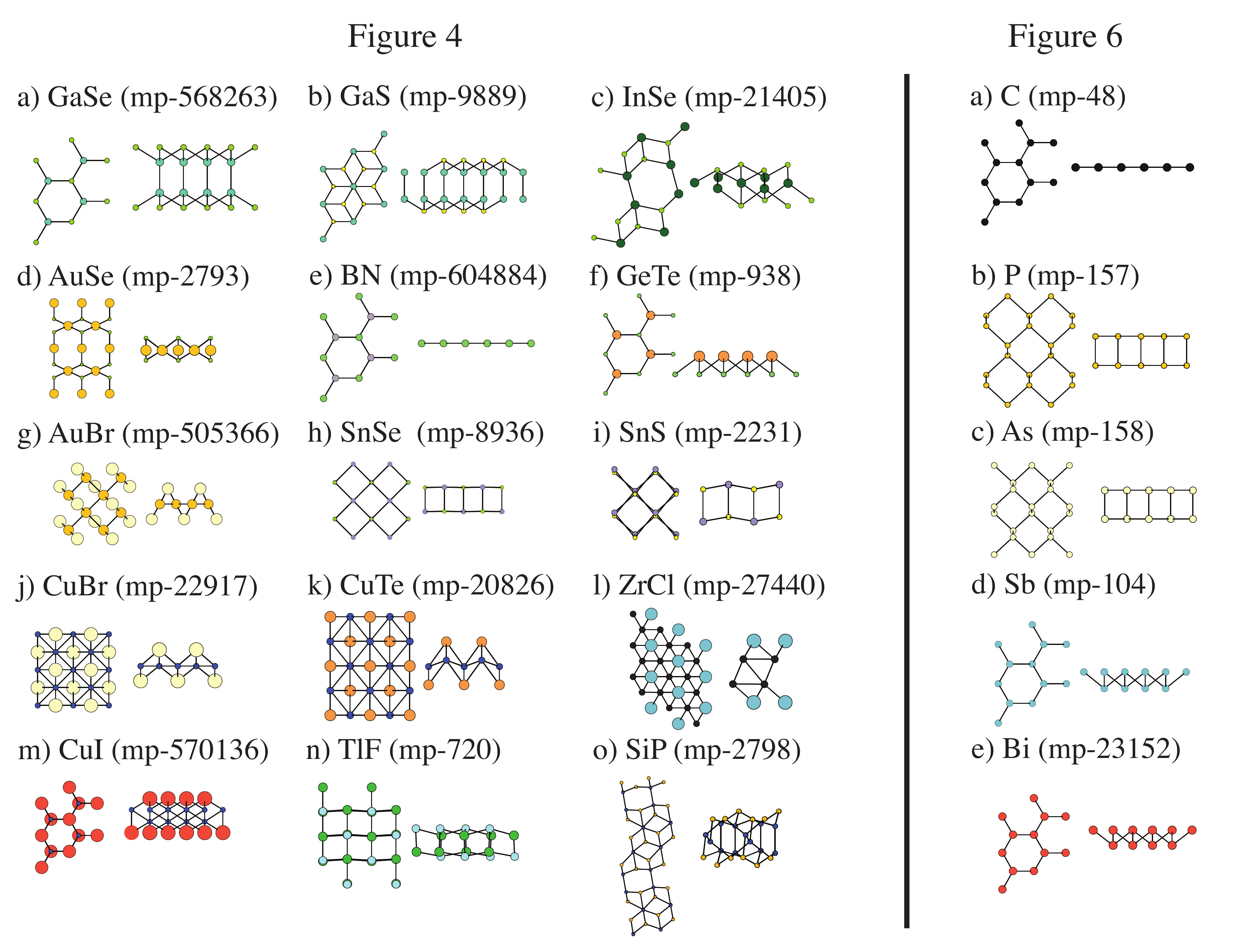}
  \caption[S1]{\label{fig:exceptions}
    (Color online) Top (left, for each material) and alternate side (right)
    views of the crystal structures shown in Figures 4 and 6 from the text. For
    the 1:1 compounds in Figure 4, the coordination numbers are largely
    controlled by the octet rule indicating covalent bonding. In structures 4(a)
    and 4(b), the Ga dimers perpendicular to the 2D layer are bonded to six
    chalcogen atoms that are either in an eclipsed or staggered configuration.
    Structure 4(e) is a planar honeycomb structure with three-fold coordinated
    atoms, a structure shared by graphene in Figure 6(a). Structure 4(f)
    presents a buckled honeycomb, where the A and B atoms are slightly displaced
    perpendicular to the plane of the 2D material. This buckled structure is
    shared by unary materials antimonene and bismene in Figures 6(d) and 6(e).
    Structures 4(j) and 4(k) are square and rectangular structures, where the
    halogen and chalcogen atoms are displaced alternately above and below the
    plane of metal atoms. In structures 4(h) and 4(i), the metal and chalcogen
    or halide atoms form distorted NaCl(100) structures consisting of two square
    checkerboard layers with varying small distortions. These structures are
    quite similar to those of phosphorene and arsenene (6(b) and 6(c)), except
    that in the unary materials the only distortions from the perfect NaCl(100)
    are in-plane.
    }
\end{figure}